# Single-particle detection of a semiconductor-to-metal transition by scanning dielectric microscopy

*Ruben Millan-Solsona[1-3], José A. Ruiz-Torres [4], Carlos Moya[5,6*], Arantxa Fraile Rodríguez[4,5], Adriana I. Figueroa[4,5], Gabriel Gomila[1,2], Amílcar Labarta[4,5], Xavier Batlle[4,5]*

[1]Nanoscale Bioelectrical Characterization Group, Institut de Bioenginyeria de Catalunya (IBEC), The Barcelona Institute of Science and Technology (BIST), Baldiri i Reixac 11-15, 08028 Barcelona, Catalonia, Spain

[2]Department d'Enginyeria Electrònica i Biomèdica, Universitat de Barcelona, Martí i Franquès, 1, 08028 Barcelona, Catalonia, Spain

[3]Present address: Center for Nanophase Materials Sciences, Oak Ridge National Laboratory, Oak Ridge, Tennessee 37831, United States of America

[4] Departament de Física de la Matèria Condensada, Universitat de Barcelona, Martí i Franquès 1, 08028 Barcelona, Catalonia, Spain

[5] Institut de Nanociència i Nanotecnologia (IN2UB), Universitat de Barcelona, Martí i Franquès, 1, 08028 Barcelona, Catalonia, Spain

[6] Departament de Química Inorgànica i Orgànica, Universitat de Barcelona, Martí i Franquès, 1, 08028 Barcelona, Catalonia, Spain

*Corresponding author: carlosmoyaalvarez@ub.edu

**Abstract**

Hybrid nanostructures that combine semiconducting and metallic components offer great potential for photothermal therapy, optoelectronics, and sensing, by integrating tunable optical properties with enhanced light absorption and charge transport. Boosting the integrated performance of these hybrid systems demands techniques capable of probing local variations of the physical properties inaccessible to bulk analysis. Here, we report the single-particle dielectric characterization of hybrid, semiconducting bismuth sulfide ($Bi_2S_3$) nanorods (NR) decorated with metallic Au nanoparticles (NP), employing scanning dielectric microscopy, which uses electrostatic force microscopy in combination with finite-element numerical simulations. We reveal a pronounced enhancement in the local dielectric response of $Bi_2S_3$ upon Au decoration, attributed to interfacial polarization and electron transfer from Au to the $Bi_2S_3$ matrix, thus suggesting a semiconductor-to-metal-like transition at the single-particle level. Numerical simulations show that the response is dominated by the vertical component of the permittivity and that the decorating metallic Au NP produce only moderate shielding of the semiconductor $Bi_2S_3$ NR core, indicating that the large increase in the dielectric response originates primarily from intrinsic modifications within the NR. Overall, these findings provide direct insight into structure-property relationships at the single-particle level, supporting the rational design of advanced hybrid nanostructures with tailored electronic functionalities.

Bismuth sulfide ($Bi_2S_3$) nanoparticles (NP) are promising candidates for optoelectronic, catalytic, and biomedical applications due to their low toxicity, narrow direct bandgap (~1.3 eV), and tunable electronic properties [1–5]. Their orthorhombic crystal structure, characterized by covalent bonding along the growth axis and weaker van der Waals forces in the perpendicular directions, favors anisotropic growth, enabling the formation of elongated morphologies such as nanorods (NR) and nanoflowers [6,7]. This intrinsic anisotropy, combined with compositional versatility, positions $Bi_2S_3$ as an excellent platform for nanoscale engineering in targeted energy conversion, imaging, and advanced catalysis.

The introduction of sulfur vacancies ($V_s$) and surface hybridization with metals are effective strategies to modulate the intrinsic properties of $Bi_2S_3$ [7–13]. These modifications generate intra-bandgap states (traps), enhance charge carrier density, and improve electronic conductivity. Au decoration amplifies these effects by forming strong Au-S bonds and donating electrons to bismuth's conduction band, which promotes additional defect states, such as bismuth antisites, and improves photothermal conversion, infrared absorption, and charge transport [9,10,12,13].

Although these modifications enhance the overall performance of $Bi_2S_3$-based materials, a deeper understanding of physical phenomena at the nanoscale can only be reached by complementing ensemble-averaged measurements and characterization techniques with single-particle experiments and methodologies, which may reveal local anisotropies, interfacial effects, and structural heterogeneities that would otherwise remain obscured in ensemble or bulk analyses. For example, recent studies have revealed magnetic polarity, switching mechanisms, and structure–property correlations in individual nanoparticles [14–17] and in plasmonic and hybrid nanostructures [18,19], providing critical input for fundamental understanding and device design.

Electrostatic force microscopy (EFM), a dynamic mode of atomic force microscopy (AFM) [20], offers a powerful approach to address this challenge. EFM enables label-free, non-contact mapping of electric polarization and dielectric response with nanometric spatial resolution [21,22]. When combined with finite-element numerical modeling of tip-sample interactions - an approach known as scanning dielectric microscopy (SDM) [23,24], it allows for quantitative determination of the dielectric constant and local conductivity of individual nanostructures [21,25–29]. These EFM-based methods have successfully revealed nanoscale electrical features often inaccessible to conventional bulk techniques in diverse systems, including inorganic dielectric nanoparticles, viruses, viral capsids, DNA, protein assemblies, organic semiconductors, and nanowires [21,22,30–33].

In this work, we perform single-particle electrical characterization of sulfur-deficient $Bi_2S_3$ NR and their Au-decorated counterparts by SDM. We quantify the effective dielectric constants ($\varepsilon_{eff}$) of individual NR and observe a substantial enhancement, from ~7-8 for $Bi_2S_3$ to values exceeding 41 in $Bi_2S_3$@Au, suggesting a semiconductor-to-metal transition. Finite-element simulations reveal that the dielectric response is dominated by the vertical component of the permittivity and that the decorating metallic NP cause only moderate shielding of the semiconductor NR core. These findings provide fundamental insight into the modulation of the dielectric behavior of semiconductor-metal nanostructures through atomic-level modifications and surface hybridization, paving the way for rational design of functional hybrid materials at the nanoscale.

Hybrid $Bi_2S_3$@Au NR designed for this study were prepared in a two-step procedure (see schematics in Fig. 1a). First, $Bi_2S_3$ NR were synthesized via a hot injection method, employing thioacetamide as the sulfur source and bismuth (III) neodecanoate as the bismuth precursor, in the presence of organic surfactants. Their composition was adjusted to be sulfur-deficient, promoting the formation of structural vacancies [34]. Second, the resulting particles were subsequently decorated with Au through a rapid room-temperature reduction process, using chloroauric acid as the Au precursor and oleylamine as both reducing and capping agent, thereby allowing the growth of Au satellite structures onto the Bi$_2$S$_3$ NR surface. Details of the materials and this experimental part are provided in Sections A.1 and A.2 in the Supporting Information, while the structural and optical properties of the samples are described in Section A.3 in the Supporting information.

Low-magnification transmission electron microscopy (TEM) characterization depicted in Fig. 1b shows Bi$_2$S$_3$ NR with an average length of 95 ± 19 nm and a width of 9 ± 3 nm, uniformly decorated with about 26 ± 5 (high-contrast) Au NP per NR, with an average diameter of 4.7 ± 0.4 nm (see histograms in Fig. BS1 in the Supporting Information). The uniformity in size and spatial arrangement of the Au NP suggest that the attachment is likely governed by lattice matching and surface energy differences between the metal and the chalcogenide (see representative images of these systems in Fig. BS2 in the Supporting Information). Besides, the selected area electron diffraction (SAED; inset to Fig. 1b) image confirms the exclusive presence of $Bi_2S_3$ and Au phases, with sharp diffraction spots indicative of high crystallinity. This is also confirmed by high-resolution (HR)-TEM characterization that reveals highly crystalline structures showing the presence of (211) planes of $Bi_2S_3$ (green) and (111) planes of Au (yellow) (see Fig. 1c), as indexed using standard diffraction files (see section A.3 in the Supporting Information). Note that these crystallographic planes - (211) for $Bi_2S_3$ and (111) for Au - are among the most compact and energetically favorable facets in their respective crystal structures, consistent with their preferential expression during growth (see section A.3 in the Supporting Information).

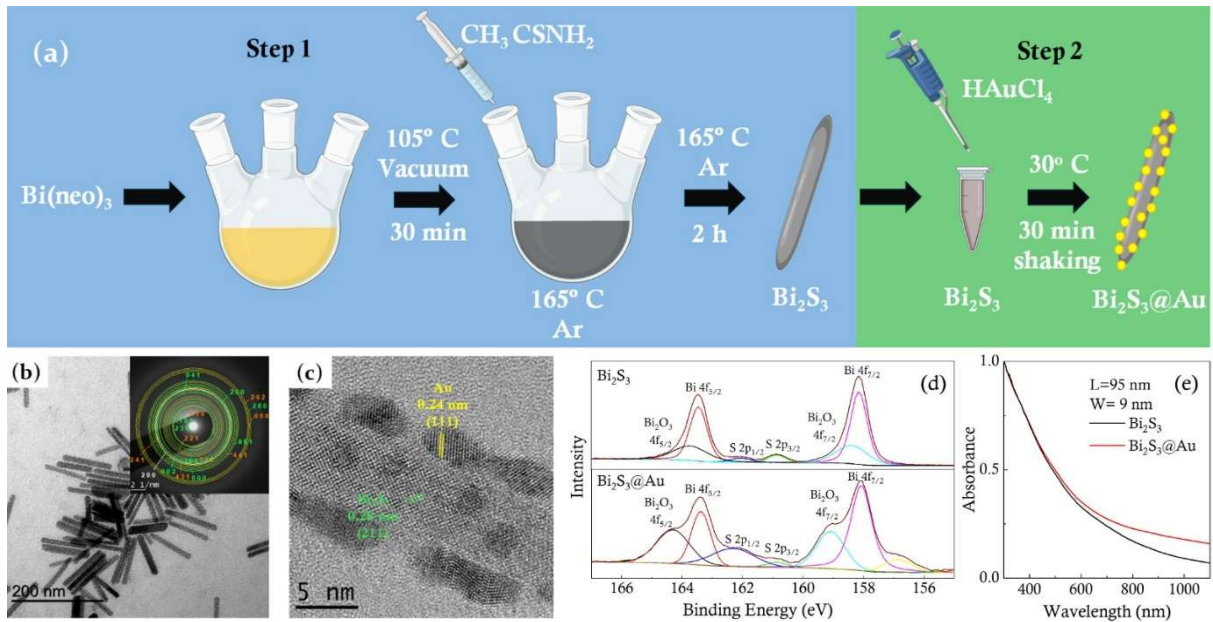

**Figure 1.** (a) Schematic of the two-step synthesis process for Bi$_2$S$_3$ and Bi$_2$S$_3$@Au NR. (b) TEM image of Bi$_2$S$_3$@Au NR with the corresponding SAED pattern (inset). (c) HRTEM image of a Bi$_2$S$_3$ NR showing Au NP attached to the surface, with lattice plane indexation. (d) XPS spectra of Bi$_2$S$_3$ (top) and Bi$_2$S$_3$@Au (bottom), highlighting the Bi 4f and S 2p core levels. (e) UV–vis absorption spectra of Bi$_2$S$_3$ NR (black) and Bi$_2$S$_3$@Au NR (red), showing enhanced near-infrared absorption following Au decoration.

X-ray photoelectron spectroscopy (XPS) was employed to investigate the surface chemical composition of these hybrid nanostructures. Figure 1d shows the high-resolution Bi 4f and S 2p core level spectra for both $Bi_2S_3$ and $Bi_2S_3$@Au samples. For $Bi_2S_3$, the Bi $4f_{7/2}$ and $4f_{5/2}$ core levels are located at 158.2 eV and 163.5 eV, respectively, while the S $2p_{3/2}$ and S $2p_{1/2}$ core levels appear at 160.9 eV and 162.1 eV, all values consistent with previous reports [8–10,35–39]. Subtle shoulders at 158.4 eV and 163.7 eV are attributed to bismuth antisites ($Bi_S$), where Bi atoms replace sulfur in the lattice [9]. Several spectral changes suggest relevant influence from Au addition. In the $Bi_2S_3$@Au sample, both Bi and S peaks shift slightly toward lower binding energies, while the $Bi_S$ signal becomes more pronounced [8–10,35–39]. This suggests a higher concentration of antisite defects after Au decoration, likely caused by sulfur depletion due to the formation of surface Au-S bonds, which favors substitution at S sites by Bi atoms. Additionally, the Au $4f_{7/2}$ and $4f_{5/2}$ core levels are shifted to higher binding energies as compared to pure, single Au NP of similar size (see Fig. BS3 in the Supporting Information) [9,12], further supporting Au-S interactions and indicating an electron transfer from Au to the $Bi_2S_3$ matrix [8–10,35–39]. The latter could be at the origin of the transition from semiconductor to metallic-like behavior demonstrated below at level of individual $Bi_2S_3$@Au particles by AFM/EFM analysis.

UV-vis spectroscopy (Fig. 1e and Fig. BS4 in the Supporting Information) was employed to investigate the optical response of $Bi_2S_3$ and $Bi_2S_3$@Au samples. There is a notable increase in the absorbance for the sulfur-deficient NR, as seen in Fig. BS4 in the Supporting Information, specially towards the near infrared (NIR) region, in agreement with previous results from the literature [8–10,35–39]. Noticeably, the Au-decorated NR exhibit enhanced absorption in the NIR region compared to the bare $Bi_2S_3$ counterparts, indicating that Au addition gives rise to distinct optical features beyond those of the semiconductor matrix. This behavior can be understood by considering the electronic structure of sulfur-deficient $Bi_2S_3$, which is expected to contain a higher density of energy states within the bandgap (traps) as compared to stoichiometric $Bi_2S_3$. These additional available states increase the density of possible electronic transitions, thereby favoring the excitation of electrons from the valence band through lower energy transitions [40,41]. Upon Au decoration, the formation of Au-S bonds promotes further sulfur depletion and Bi substitution, resulting in an even greater density of bandgap, trap-related states. Moreover, since the Fermi level of Au lies within the bandgap of $Bi_2S_3$, electron deexcitation from the conduction band of the $Bi_2S_3$ matrix to Au conduction band can also enhance nonradiative recombination, contributing to the increased infrared absorption found in $Bi_2S_3$@Au [8–10,35–39].

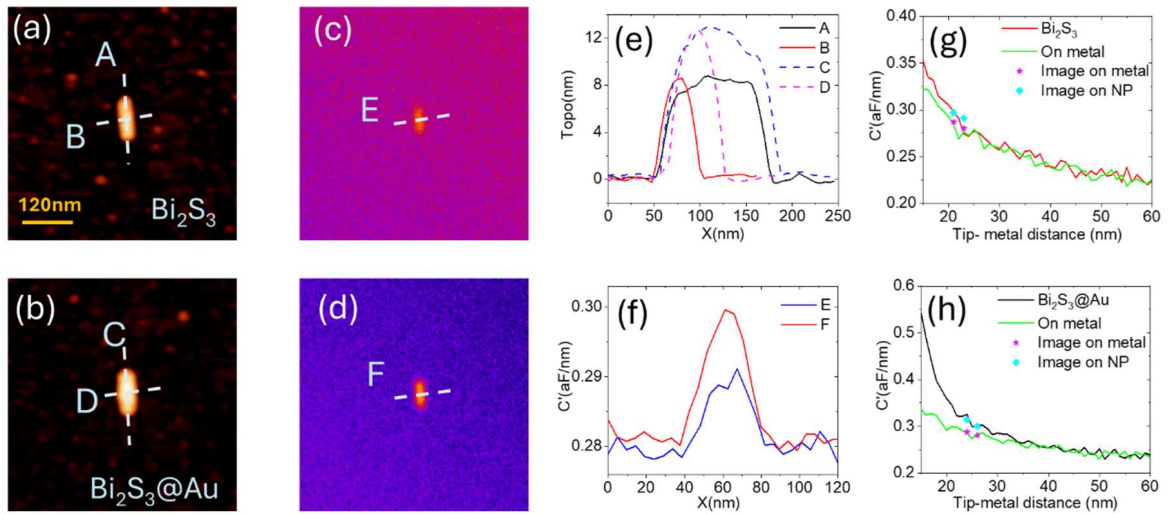

**Figure 2.** Nanoscale topographic and dielectric characterization of individual $Bi_2S_3$ and $Bi_2S_3$@Au NR. (a), (b) AFM topographic images of $Bi_2S_3$ and $Bi_2S_3$@Au NR, respectively. (c), (d) EFM images acquired in constant height mode at tip-metal substrate distances of $21 \pm 2$ nm for $Bi_2S_3$ and $26 \pm 2$ nm for $Bi_2S_3$@Au. (e) Corresponding topographic profiles along the dashed lines indicated in (a) and (b): A and B (solid black and red lines, respectively) for a $Bi_2S_3$ NR, and C and D (dashed blue and violet lines, respectively) for a $Bi_2S_3$@Au NR. (f) Capacitance gradient profiles (solid blue and red curves) obtained from the scans indicated by dashed lines (E and F) in panels (c) and (d), respectively. (g), (h) Electrical approach curves measured over the clean metallic substrate and the $Bi_2S_3$ and $Bi_2S_3$@Au NRs, respectively.

Samples for EFM were prepared by drop-casting hexane solutions of dilute NR onto flat Au substrates and evaporation under controlled vacuum conditions (see details in Section A.4. in the Supporting Information). EFM images were collected in constant height mode on single NR with an approximate tip-metal substrate distance of 25 nm. Figure 2 summarizes the nanoscale topographic and dielectric characterization of single $Bi_2S_3$ and $Bi_2S_3$@Au NR, including topographic and capacitance gradient ($C'$) images, height profiles, and electrical approach curves onto the metallic surface and onto the NRs (see details in Sections A.5. and A.6. in the Supporting Information). At first glance, the Au-coated $Bi_2S_3$ NR appears slightly longer than the uncoated NR (see Fig. 2a,b), as shown by the height profiles in Fig. 2e. The $Bi_2S_3$ NR exhibits maximum sizes (width × length) of 9 × 75 nm (profiles A and B), while the $Bi_2S_3$@Au NR reaches 13 × 82 nm (profiles C and D). These sizes were obtained considering the tip–sample convolution effect (see details in Section A.7 and Fig. BS5 in the Supporting Information).

Next, the effect of Au decoration on the dielectric response of individual NR was investigated by EFM, as shown in Fig. 2c,d for $Bi_2S_3$ and $Bi_2S_3$@Au NR, respectively. As a common trend, both samples exhibit a dielectric constant significantly higher than that of air ($\varepsilon_{eff} > 1$), as indicated by the enhanced signal in the $C'$ images when the tip is positioned over the NR compared to the substrate. Furthermore, there is a higher dielectric response exhibited by the $Bi_2S_3$@Au NR, as the $C'$ profile (profile F in Fig. 2f) shows a slightly higher peak (~0.295 aF/nm) compared to that corresponding to the $Bi_2S_3$ NR (~0.288 aF/nm, profile E), suggesting that Au decoration increases the local polarizability of the NR, i.e., $\varepsilon_{Bi_2S_3} < \varepsilon_{Bi_2S_3@Au}$. Note that this comparison was made under similar tip-sample distances. Specifically, based on the signal measured over the metallic substrate in the image, the tip-sample distances were estimated to be approximately 21 ± 2 nm and 26 ± 2 nm for the $Bi_2S_3$ and $Bi_2S_3$@Au NR, respectively. This enhancement of the dielectric response is further supported by the electrical approach curves acquired under similar conditions (Fig. 2g,h), where $C'$ reaches ~0.33 aF/nm for $Bi_2S_3$ and ~0.53 aF/nm for Bi$_2$S$_3$@Au NR at tip-metal substrate distances of around 15 nm. Additional results in agreement with these general trends are shown in Fig. BS6 in the Supporting Information for two other NR.

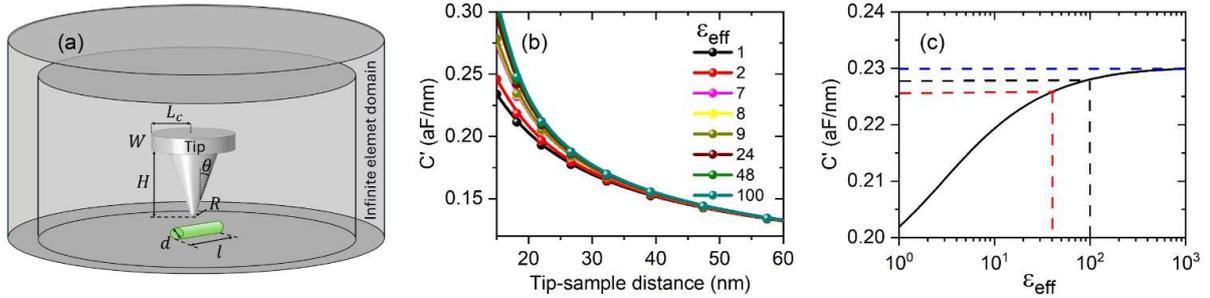

**Figure 3.** (a) Finite-element model used to calculate $\varepsilon_{eff}$ of the NR, consisting of a tip above a cylindrical NR (length $l$, diameter $d$) on a grounded metallic substrate, enclosed in a cylindrical domain with infinite boundary conditions (not to scale). Simulation parameters: tip radius R = 35 nm, half-cone angle θ = 22°, cone height H = 12.5 μm, disk width W = 3 μm, disk radius $L_c$ = 8 μm; cylinder diameter = 10 nm, and cylinder length = 80 nm. (b) Simulated $C'(z)$ approach curves for several values of $\varepsilon_{eff}$, where z stands for the tip-substrate distance, showing superimposing curves for $\varepsilon_{eff}$ = 48 and 100, thus setting an upper limit for the highest measurable value of $\varepsilon_{eff}$. (c) $C'$ values as a function of $\varepsilon_{eff}$ at a fixed tip–substrate distance of 20 nm, tending to an asymptotic value of ~0.23 aF/nm (blue dashed line). The red dashed line indicates the lower uncertainty bound in the experimental EFM data taken from the asymptotic value.

While the previous results based on $C'$ provide qualitative insight into the dielectric properties of the NR, a quantitative analysis requires a more rigorous approach to calculate reliable values of the $\varepsilon_{eff}$ of the NR. To this end, the geometrical and physical features of the system must be carefully incorporated into a 3D finite-element model. Figure 3 depicts the results of the simulations of the model, including a scheme of the tip-NR-substrate geometry (Fig. 3a), simulated approach curves for several values of $\varepsilon_{eff}$ (Fig. 3b), and the resulting $C'(\varepsilon_{eff})$ curve at a fixed tip-substrate distance of $z = 20$ nm (Fig. 3c).

For accurate modeling, a first calibration of the tip geometry was required by recording experimental approach curves acquired on a clean metallic region of the substrate. These curves were fitted to numerical simulations of a tip-on-metal system to obtain key geometric parameters. Specifically, the tip radius was determined to be R = 35 ± 2 nm, which is in good agreement with the deconvolution analysis from AFM topography (R = 37 ± 4 nm), and the half-cone angle was set to a nominal value of θ = 22°. This calibration, described in Section A.7 in the Supporting Information, ensures that the tip geometry used in the simulations matches the real experimental conditions, thus allowing reliable values of $\varepsilon_{eff}$ to be determined from the EFM data.

Using this calibrated geometry and the 3D tip-cylinder model, $\varepsilon_{eff}$ of individual NR was calculated by fitting simulated $C'(z)$ approach curves to the experimental ones (see Section A.7 and Fig. BS7 in the Supporting Information). It is worth mentioning that quantitative interpretation of this result is limited by the loss of sensitivity in the high-permittivity range. As shown in Fig. 3b, the simulated approach curves for cylindrical objects with $\varepsilon_{eff}$ ranging from 1 to 100 superimpose for $\varepsilon_{eff} = 48$ and 100, reflecting the onset of saturation. This saturation is further illustrated in Fig. 3c, which displays simulated values of $C'(\varepsilon_{eff})$ at a fixed tip–substrate distance of $z = 20$ nm. At high permittivity, $C'$ tends to an asymptotic value of ~0.23 aF/nm (blue dashed line in Fig. 3c), indicating that further increases in $\varepsilon_{eff}$ above a certain limiting value cannot be reliably resolved due to the uncertainty in the experimental EFM data. This uncertainty extends approximately from the asymptotic value of $C'$ to the red dashed line in Fig. 3c. So, one can estimate a limiting value of $\varepsilon_{eff} = 41 \pm 2$ from the intersection between the red dashed line and $C'$. Any permittivity above this threshold would yield simulated curves consistent with the experimental data owing to saturation and the uncertainty range. By applying the fitting procedure described above, the obtained values of $\varepsilon_{eff}$ for the two Bi$_2$S$_3$ NR studied in this work (see Figs. 2g and BS6g in the Supporting Information for the corresponding approach curves) were $8 \pm 1$ and $7 \pm 1$, while for the two Bi$_2$S$_3$@Au NR (Figs. 2h and Fig. BS6h in the Supporting Information) significantly higher values around 100 were obtained, suggesting a transition towards metallic-like behavior. While those fitted values should be considered indeterminate within the range 41-100 due to experimental uncertainty in the EFM data, which sets an upper confidence limit, the large increase in the effective permittivity of the hybrid NR supports the hypothesis that Au decoration enhances local electronic polarizability. This is consistent with partial metallic behavior likely originating from interfacial charge transfer between Au NP and the Bi$_2$S$_3$ NR, and/or direct shielding effects induced by the Au NP.

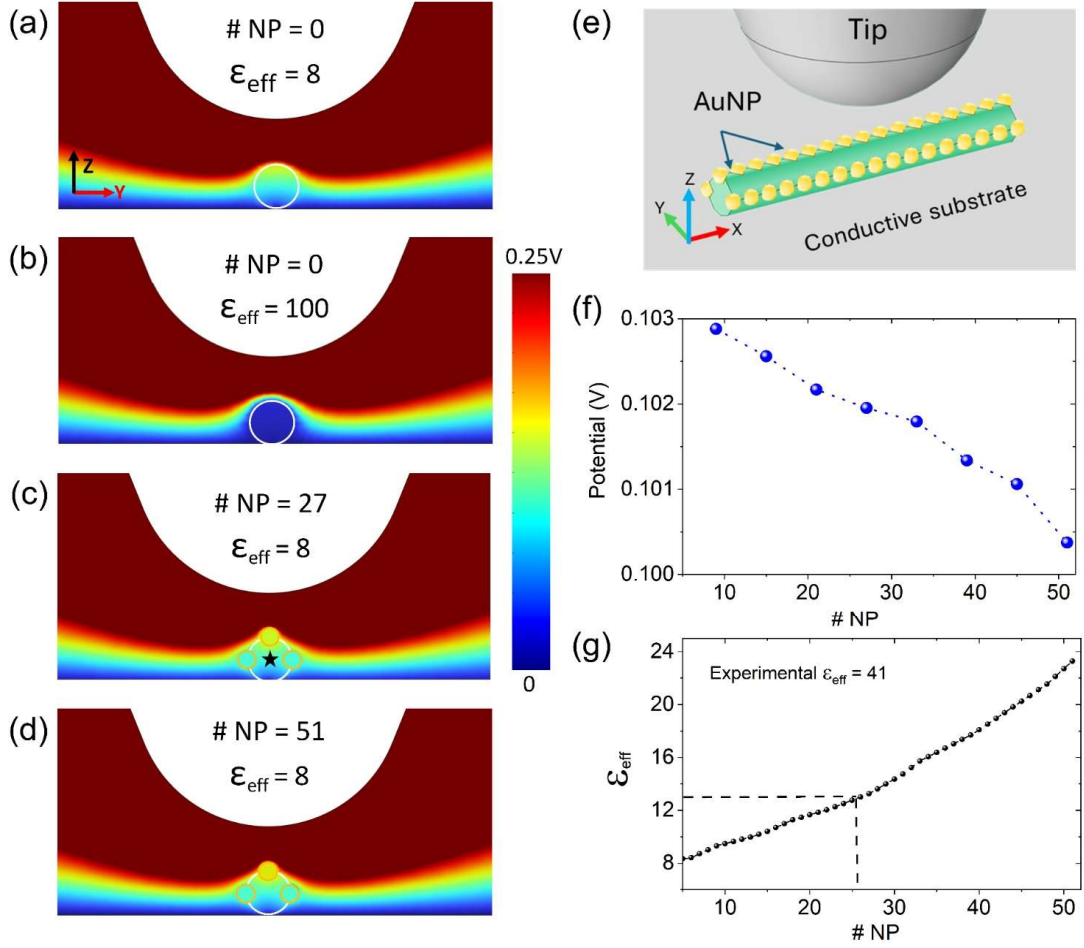

**Figure 4**. (a)–(d) Simulated electrostatic potential maps in the y–z plane across the nanorod for different effective dielectric constants and varying numbers of Au NP attached to the surface. Color scale indicates the correspondence between colors and the values of the electrostatic potential. (e) Schematic of the 3D model showing Au NP arranged along the NR surface beneath the tip. (f) Electrostatic potential at the center of the NR (point marked with a star in (c)) as a function of the number of Au NP for $\varepsilon_{eff} = 8$. The electrostatic potential decreases with increasing number of Au NP, as expected for a partially screened system, but without reaching complete shielding of the NR interior, even for about twice the number of Au NP attached on the hybrid NR. (g) Effective permittivity as a function of the number of Au NP. In the experimental case, we estimate an average of ~26 NP per NR, corresponding to an effective permittivity of about 13 for the simulated. curve in pannel (g), which is significantly lower than the value within 41-100 obtained from EFM curves for Bi$_2$S$_3$@Au NR.

A key question is whether the Au NP on the surface of Bi$_2$S$_3$@Au NR could mask their internal response, potentially leading to a misinterpretation of the observed dielectric enhancement. To address this issue, surface-attached Au NP were included in the model described in Fig. 3 to quantify how their presence contributes to the enhancement of the measured $\varepsilon_{eff}$. The results of these simulations are shown in Fig. 4. Figure 4a–d shows the simulated electrostatic potential in the $y-z$ plane for $\varepsilon_{eff} = 8\ and\ 100$, as well as for increasing numbers of Au NP decorating the $Bi_2S_3$ NR. Figure 4e illustrates the geometry of the attached Au NP, which are arranged along three equidistant longitudinal rows. Figure 4a,b compares the cases of bare NR with $\varepsilon_{eff} = 8$ and 100, showing that an enhancement of the dielectric constant of the order of magnitude of that observed experimentally results in almost complete shielding of the NR interior (the electrostatic potential distribution is negligible inside the NR) as expected for metal-like behavior. In contrast, keeping $\varepsilon_{eff} = 8$ for the NR and attaching an increasing number of Au NP up to the maximum allowed by the selected geometry (51) only reveals moderate changes in the potential distribution, associated with partial shielding of the NR interior (see Fig. 4c,d). It is worth remarking, that the average number of Au NP on the $Bi_2S_3$@Au NR is about 26 ± 5 (see Fig. BS1d). To quantify this shielding effect, the electrostatic potential at the center of the NR as a function of the number of Au NP is shown in Fig. 4f, exhibiting a slight monotonous decrease of about 4%. Finally, Fig. 4g shows the effective permittivity of the hybrid structure as a function of the number of Au NP, from which a value of $\varepsilon_{eff} \approx 13$ for 26 NPs is estimated. This value is significantly lower than the 41-100 range found experimentally for $Bi_2S_3$@Au NR. Furthermore, additional simulations with an anisotropic NR (see section A.10 and Fig. SB8 in the Supporting Information) show that the experimentally observed dielectric enhancement could be strongly anisotropic and mainly due to the vertical component of the permittivity.

In summary, $Bi_2S_3$ NR exhibit a clear transition from semiconductor to metallic-like behavior upon surface decoration with Au NP, shown by EFM combined with finite-element simulations. We quantified the dielectric response of individual NR and demonstrated that Au decoration significantly enhances their polarizability, likely driven by interfacial charge transfer and the formation of bandgap trap states related to sulfur deficiency and Bi antisites. The observed dielectric enhancement primarily reflects the vertical component of the permittivity, as the technique is sensitive only to this direction and does not capture the full anisotropy of the material. Simulations show only moderate shielding of the NR core by the surface Au NP, supporting that the enhanced dielectric response originates from intrinsic modifications within the NR. These results offer direct evidence of a semiconductor-to-metal-like transition at the single-particle level, often masked in bulk measurements. This approach offers a powerful framework to probe local electronic behavior and dielectric modulation in individual NR. Its extension to other hybrid systems could drive the rational design of nanomaterials with tunable optoelectronic and photothermal functionalities.

**Author Contributions**

XB, AL, RM, GG and AIF conceived the research. CM and JART carried out the synthesis, the structural, compositional, and optical characterization, together with XB. RM conducted the AFM experiments, analysis of EFM data and carried out numerical simulations. AL, AFR, CM, XB, GG, and AIF contributed to the discussion and interpretation of the characterization results. CM, JART, RM, and AL led the manuscript writing, with input from all authors. All authors reviewed and approved the final version of the manuscript.

**Acknowledgement**

The authors gratefully acknowledge financial support from the Spanish MICIIN (grant numbers PGC2018-097789-B-I00, PID2021-127397NB-I00, PID2022-142297NB-I00), Catalan AGAUR (Groups of Excellence 2021SGR00328 and 2021SGR00453),

University of Barcelona (IN2UB-ART2022 project), and the European Union FEDER funds. Most of the characterization of the samples was performed at the scientific facilities of the University of Barcelona (CCiTUB), with special thanks to the Surface Analysis Laboratory. We thank E. Carranza-Botey, J. Diago-Forero, S. Hernandez and B. Garrido for their collaboration in the optical characterization of the samples and for fruitful discussions. CM acknowledges funding from the University of Barcelona and the Spanish Ministry of Universities under the Maria Zambrano Program, funded by the European Union Next Generation EU/PRTR, as well as the Beatriu de Pinós fellowship program (2022 BP 00243). AIF and CM are Serra Húnter fellows. JR acknowledges funding from PRE2022-102644 financed by MICIU/AEI/10.13039/501100011033 and FSE+. GG acknowledges funding from the ICREA Academia Program from the Generalitat de Catalunya.

# Supporting Information for

# Single-particle detection of a semiconductor-to-metal transition by scanning dielectric microscopy

*Ruben Millan-Solsona[1-3], José A. Ruiz-Torres [4], Carlos Moya[5,6*], Arantxa Fraile Rodríguez[4,5], Adriana I. Figueroa[4,5], Gabriel Gomila[1,2], Amílcar Labarta[4,5], Xavier Batlle[4,5]*

[1]Nanoscale Bioelectrical Characterization Group, Institut de Bioenginyeria de Catalunya (IBEC), The Barcelona Institute of Science and Technology (BIST), Baldiri i Reixac 11-15, 08028 Barcelona, Catalonia, Spain

[2]Department d'Enginyeria Electrònica i Biomèdica, Universitat de Barcelona, Martí i Franquès, 1, 08028 Barcelona, Catalonia, Spain

[3]Present address: Center for Nanophase Materials Sciences, Oak Ridge National Laboratory, Oak Ridge, Tennessee 37831, United States of America

[4] Departament de Física de la Matèria Condensada, Universitat de Barcelona, Martí i Franquès 1, 08028 Barcelona, Catalonia, Spain

[5] Institut de Nanociència i Nanotecnologia (IN2UB), Universitat de Barcelona, Martí i Franquès, 1, 08028 Barcelona, Catalonia, Spain

[6] Departament de Química Inorgànica i Orgànica, Universitat de Barcelona, Martí i Franquès, 1, 08028 Barcelona, Catalonia, Spain

*Corresponding author: carlosmoyaalvarez@ub.edu

**Contents**

**A. Methods**

**B. Supporting Figures**

**C. References**

## A. Methods

### A. 1. Materials

Bismuth (III) neodecanoate (Bi(neo)$_3$), oleic acid (90%), thioacetamide (≥ 99%), oleylamine (70%), 1-octadecene (90%), and 2-Propanol (99.9%), acquired from Sigma-Aldrich. Chloroauric acid trihydrate (HAuCl$_4$, ≥ 49.0%, Au) was purchased from Acros Organics. Nitric acid (HNO$_3$, 65%), hydrochloric acid (HCl, 37%), acetone (99,5%), hexane (95%), and toluene (99.5%) were bought from Panreac. All the substances were used as received without further purification, although thioacetamide lumps were grounded before use.

### A.2. Synthesis of Bi$_2$S$_3$@Au NR

Bi$_2$S$_3$@Au NR were prepared following a two-step synthetic route as follows. First, Bi$_2$S$_3$ NR were synthesized following our previous protocol with modifications to enhance particle dimensions [1]. Briefly, 0.076 g of thioacetamide (1 mmol) and 0.374 g of oleylamine (1.5 mmol) were combined with 0.5 mL of 1-octadecene in a small vial. This mixture was sonicated at 60°C for 45 min until a homogeneous yellow solution formed. Meanwhile, 0.723 g of Bi(neo)$_3$ (1 mmol) and 3.64 g of oleic acid (12 mmol) were dispersed in 5 mL of 1-octadecene in a three-neck round-bottom flask with a Teflon-coated magnetic stir bar spinning at 700 rpm. The reaction mixture was purged under low pressure (P ≤ 100 Pa) at 165°C for 30 min using a Schlenk line. After purging, the Schlenk line was switched to an argon flow, and the pale-yellow mixture was heated to 165°C at a rate of 6°C/min and maintained at this temperature for 20 min to homogenize the mixture. Subsequently, the flask was cooled to 105°C at a rate of 2°C/min, and the thioacetamide solution was rapidly injected into the reaction mixture. The solution's color immediately changed from pale yellow to black, and the mixture was maintained at this temperature for 120 min. The reaction mixture was then cooled to room temperature by removing the heating mantle and washed by centrifugation twice using acetone as an antisolvent. The final pellet was dried with compressed air, redispersed in 5 mL of toluene, and stored in the fridge at 5ºC.

Second, Au NP were attached to the $Bi_2S_3$ surface by a rapid reduction process in presence of oleylamine. A solution of 0.148 g (0.25 mmol) of $HAuCl_4$ and 0.4g (1.6 mmol) of oleylamine was prepared using 1.5 mL of toluene as the solvent. This mixture was sonicated for 10 min, resulting in a homogeneous yellow solution. Then, 140 µL of the Au solution was added to a toluene of NR solution containing 1 mg Bi/mL, 2.2 mM) in a total volume of 1 mL. After shaking the mixture at 1000 rpm and 30°C for 30 min, the sample was centrifuged at 1000 rpm for 10 min. Finally, the supernatant was discarded, and the pellet was redispersed in 1 mL of toluene and stored at 5°C.

**A.3. Structural and optical characterization**

Transmission electron microscopy (TEM). Samples prepared in section A.2 were studied using a JEOL 1010 microscope operating at 80 kV. The samples were prepared by depositing one drop of the diluted suspension in toluene on a carbon-coated Cu grid and left dried at room temperature. For each sample, the length and width of at least 300 particles were analyzed using the Image J software [2]. The obtained histograms of the width W and the length L of the NRs were then fitted to log-normal distributions of the general form:

$$P(D) = \frac{1}{S\sqrt{2\pi}D} e^{-\ln^2\left(\frac{D}{D_o}\right)/(2S^2)} \qquad (1)$$

where $D$ stands for the specific quantity to be studied, $D_o$ is the corresponding most probable value, and $S$ is the standard deviation of the logarithmic distribution. The mean particle size $D_{TEM}$ and standard deviation $\sigma$ were computed using the following equations

$$D_{TEM} = D_o e^{-S^2/2} \qquad (2)$$

$$\sigma = D_o e^{-S^2/2}\sqrt{e^{-S^2} - 1} \qquad (3)$$

and the relative standard deviation $RSD = \sigma/D_{TEM}$ was defined to compare the dispersion of the NPs' size.

High-Resolution TEM (HRTEM). The crystal structure of samples was studied using a JEOL 2010 microscope operating at 200 kV. $Bi_2S_3$ NR were indexed to (hkl) planes using reference pattern 01-075-1306, while the Au NPs were assigned to crystallographic planes based on reference 03-065-2870, using Gatan's Digital Micrograph software.

X-ray photoelectron spectroscopy (XPS). XPS experiments were carried out in ESFOSCAN at the CCiTUB, an equipment based on the PHI VersaProbe 4 instrument from Physical Electronics. Measurements were carried out with a monochromatic focused X-ray source (Aluminium K-alfa line of 1486.6 eV) calibrated using the $3d_{5/2}$ line of Ag with a full width at half maximum of 0.6 eV. The area analyzed was a circle of 100 μm in diameter, and the resolution selected for the spectra was 224 eV Pass Energy and 0.8 eV/step for the general spectra and 27 eV Pass Energy and 0.1 eV/ step for high resolution spectra of the selected elements. All measurements were performed in an ultra-high vacuum chamber at a pressure between $5\times10^{-10}$ and $5\times10^{-9}$ Torr. The background signal was calculated using a Shirley routine [3], which assumes an S-shaped background per peak. Then, the peaks were fitted to Gaussian-Lorentzian curves using a Multiple peak fit tool.

Ultraviolet-visible (UV-vis) spectroscopy. UV-vis measurements were carried out using an Agilent Cary 3500 ultraviolet-visible spectrometer ranging from 200 to 1100 nm with a spectral resolution of 0.5 nm and using 3.5 mL Suprasil® quartz cuvettes with a 10 mm pathlength.

Inductively coupled plasma optical emission spectroscopy (ICP-OES). Aliquots of sample solutions were dried and then digested in 400 μL of aqua regia under sonication for 30 min Then, the samples were diluted in volumetric flasks with distilled water. The element concentration was determined with a Perkin Elmer OPTIMA 3200RL at Serveis Cientifico-Tècnics of the Universitat de Barcelona (CCiTUB).

### A.4. Sample preparation for AFM/EFM imaging

Substrates for these experiments consisted of atomically flat gold films on mica. Before particle deposition, substrates were cleaned by ultraviolet ozone treatment for 10 min. Next, the sample solutions with Bi concentrations of 0.05 mg/mL were sonicated for five min. A 15 μL drop was then deposited onto the substrates and allowed to dry in a vacuum chamber for 1 h and cleaned with acetone and 2-propanol, then dried completely in a vacuum chamber.

### A.5. Atomic force microscopy imaging

We used a Cypher S AFM from Oxford Technologies. AFM topographic images of 128 x 128 pixels (5 x 5 μm² and 0.6 x 0.6 μm²) were obtained at 0.5 Hz per line in intermittent contact mode using PtSi-CONT conductive probes from Nanosensors with a spring constant $k \sim 0.3$ N/m (as determined by the provider based on the probe dimensions), resonance frequency $fr \sim 13$ kHz, nominal tip radius $R \sim 20$ nm, and half-cone angle $\theta \sim 20°$. AFM imaging was performed according to the following protocol: initial $5 \times 5$ μm² scans were used to locate isolated particles, followed by high-resolution $0.6 \times 0.6$ μm² scans targeting individual ones. Image processing was conducted using WSxM software [4] (Nanotec Electrónica S.L.) and custom MATLAB scripts.

### A.6. Electrostatic force microscopy imaging

The usual model for electrostatic force in EFM is based on the expression of capacitor energy, $W(t) = \frac{1}{2}CV(t)^2$, where a time-dependent potential difference is applied to the two electrodes, represented by the conductive tip and the metallic substrate of the sample. In the case of periodic potentials of the form $V(t) = V_{DC} + V_{AC}\cos(\omega t)$, the electrostatic force is calculated by deriving with respect to the tip-substrate distance $z$ and separating by the different frequency components, such that the resulting force is

$$F_z(t) = F_{DC} + F_\omega \cos(\omega t) + F_{2\omega} \cos(2\omega t) \qquad (4)$$

The capacitance $C$ is assumed to be independent of frequency. Of these three components, the second harmonic component $2\omega$ is the only one that does not depend on the static voltages and is used for capacitance measurements in EFM. Thus, the second harmonic oscillation amplitude is given by

$$F_{2\omega} = \frac{1}{4}\frac{\partial C}{\partial z} V_{AC}^2 \qquad (5)$$

and from this expression, the capacitance gradient is determined as a function of the oscillation amplitude and the applied AC voltage.

EFM images were obtained using the AFM system and probes described above. The $2\omega$ harmonic of the probe oscillation amplitude, $F_{2\omega}$, was recorded using an internal lock-in amplifier of the Cypher in SNAP two-pass mode, a line-by-line constant height mode built into the AFM system, as detailed in previous works [5–8]. EFM data were acquired with a voltage amplitude $V_{ac}$ = 5 V at a frequency $f$ = 2 kHz. The instrumental noise of the capacitance gradient was in the range of 2 zF/nm. EFM measurements were conducted in controlled dry air conditions (RH <1%) maintained by an N$_2$ flow.

### A.7. Tip geometry calibration and determination of the effective dielectric constant

The tip radius $R$, half-cone angle $\theta$, and capacitance gradient offset $C'_{offset}$ used in the theoretical model were determined as detailed elsewhere [6,7,9,10]. Basically, the approach curves of the capacitance gradient $C'(z)$ for a tip on metal model are numerically calculated and the experimental curves $C'(z)$ measured in a bare region of the substrate are fitted. In the analysis, the microscopic parts of the tip were adjusted to their nominal values $\theta$ = 22º, $H$ =

12.5 μm, $W$ = 3 μm, and $Lc$ = 8 μm. The effective dielectric constant of the NRs was determined by least-squares fitting of the experimental approach curves $C'(z)$ measured approximately at the center of the NR with those numerically calculated using the model described in Fig. 3a (see Fig. BS7).

### A.8. Determination of NR geometry from AFM profiles

For simplicity, the NR was modeled as a cylinder with diameter $d$ and length $l$, allowing determination of the NR diameter from the height of the vertical and longitudinal profile of the flattened topographic images using WSxM software. To determine the length of the NR, a simple deconvolution model was used (see Fig. BS5a,b), allowing for an analytical relationship between the tip radius, NR height, and the width of the experimental profile. In our case, we obtained a tip radius of 37 ± 4 nm, as shown in the calculation example in Fig. BS5.

### A.9. Finite-element numerical simulations

Quantitative analysis of the data acquired by EFM was performed following the methods of Scanning Dielectric Microscopy [11,12], adapting finite element models to the specific geometry for modeling the NRs. In the quantitative analysis, the model described in Fig. 3a was used for $Bi_2S_3$ NRs, and the model in Fig. 4e was used for $Bi_2S_3$@Au NR. In both models, the tip is modeled as usual, using a truncated cone with height $H$ and half-angle $\theta$, ending in a tangent sphere with radius $R$. On top of the cone, there is a "cantilever" disk with thickness $W$ and radius $Lc$ to model local cantilever effects. The lever portion of the probe was not explicitly modeled, and its effects were included by a phenomenological capacitance gradient offset, $C'_{offset}$. The $Bi_2S_3$ NRs were modeled as a cylinder with diameter $d$ and length $l$ with a dielectric constant $\varepsilon_{eff}$ on a conductive substrate. For $Bi_2S_3$@Au NR, spheres with a very high dielectric constant were added to the cylinder surface to model the metallic behavior of the gold NP. In the case of studying the anisotropic dielectric constant of the cylinder, a diagonal relative permittivity of the form

$$\varepsilon_{eff} = \begin{pmatrix} \varepsilon_\parallel & 0 & 0 \\ 0 & \varepsilon_\parallel & 0 \\ 0 & 0 & \varepsilon_z \end{pmatrix} \tag{6}$$

was considered. The electrostatic force acting on the tip was determined by solving Poisson's equation for the described model and integrating Maxwell's stress tensor on the tip surface, using the electrostatic module of Comsol Multiphysics 6.0 and custom codes written in Matlab (The Mathworks) as detailed in previous works [6,7,9,10].

**A.10. Asymmetric effective dielectric constant**

To investigate the influence of asymmetric quantum confinement on the properties of the NR [13–16], we modeled their dielectric response as anisotropic, distinguishing between components along and perpendicular to the rod axis. Thus, we separated the dielectric constant into two components: one along the longitudinal axis ($\varepsilon_\parallel$) and one perpendicular to it ($\varepsilon_\perp = \varepsilon_z$), assigned to the $z$-direction in this case. Simulations incorporating this anisotropy were used to calculate the electrostatic potential distribution in the $y - z$ plane (Fig. BS8a–d), based on the model illustrated in Fig. BS8e. Each dielectric component was independently varied across a wide range—from low (dielectric-like) to high (metallic-like) values. The results reveal that when $\varepsilon_z$ is high, the NR potential closely approaches zero, consistent with metallic behavior. In contrast, a high $\varepsilon_\parallel$ alone yields a more gradual potential distribution, characteristic of a dielectric. This indicates that the parallel component $\varepsilon_\parallel$ has limited impact on the potential profile and, therefore, on the electrostatic force experienced by the tip.

Figure BS8f shows the simulated capacitance gradient $C'$ as a function of $\varepsilon_z$ for various $\varepsilon_\parallel$ values, using a fixed tip–substrate distance of 20 nm. The influence of $\varepsilon_\parallel$ becomes noticeable

only when $\varepsilon_z$ is already high, and even then, the variation in $C'$ does not exceed ~4 zF/nm. This confirms that the measurement is primarily sensitive to $\varepsilon_z$. Applying the same analytical framework as before and considering the experimental uncertainty, two limiting cases emerge: if $\varepsilon_\parallel$ is assumed to be very high, the lower bound for $\varepsilon_z$ is refined to be $\varepsilon_z > 28 \pm 2$ (blue arrow in Fig. BS8f). Conversely, if $\varepsilon_\parallel$ is low, the condition $\varepsilon_z > 100$ is required to match the experimental data.

## B. Supporting Figures

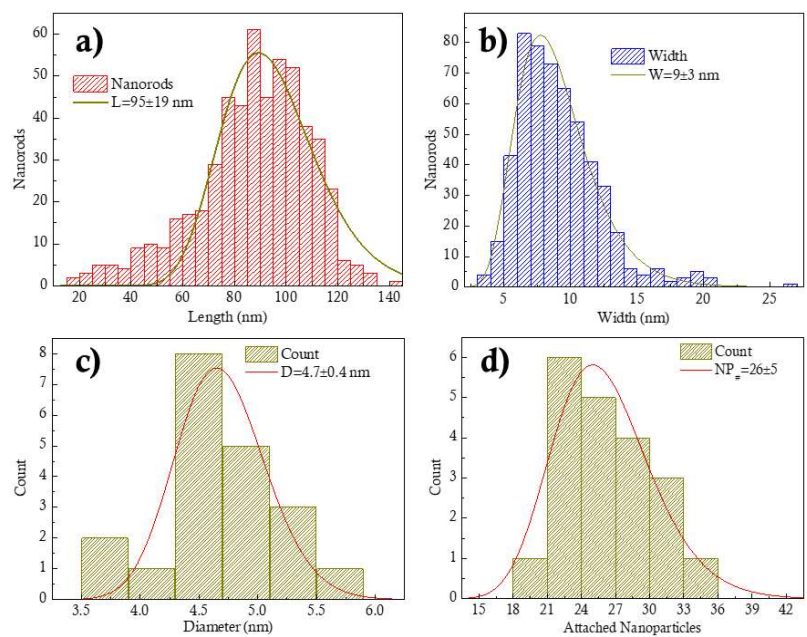

**Figure BS1.** Histograms of $Bi_2S_3$@Au of (a) length, (b) width, (c) Au NP diameter, and (d) number of attached Au NP per NR.

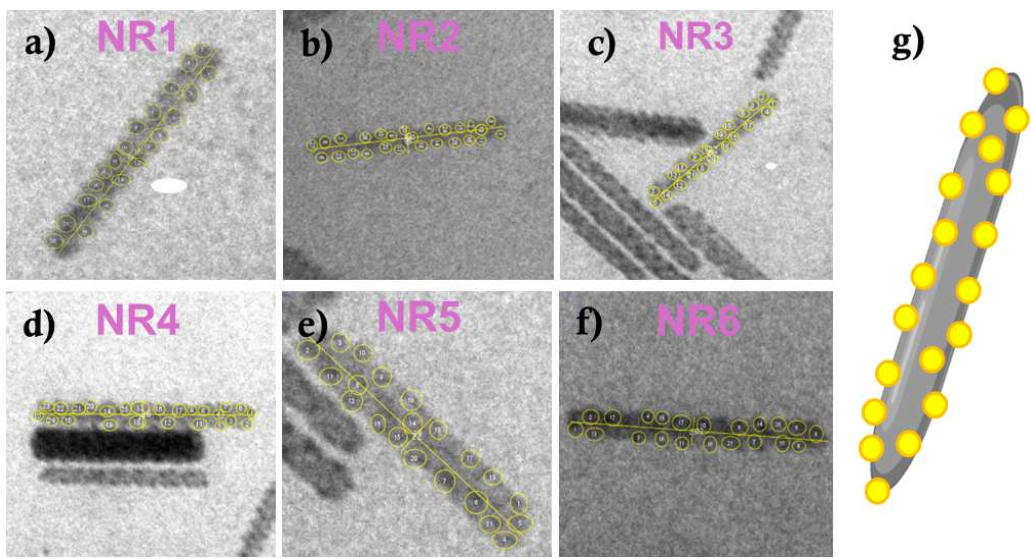

**Figure BS2.** TEM images of six representative Bi$_2$S$_3$@Au NR (a)–(f), labeled NR1–NR6, with Au NPs highlighted by yellow circles. (g) Schematic illustration of a Bi$_2$S$_3$ NR decorated with Au NPs along its surface.

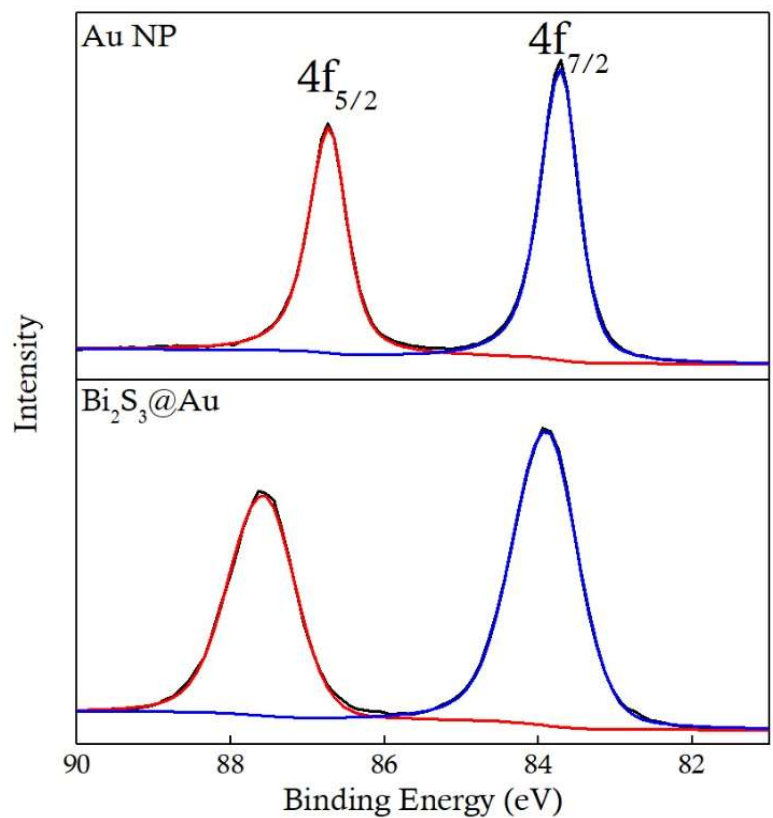

**Figure BS3.** Comparison of high-resolution XPS spectra of the Au 4f core level. Top: Spectrum of pure Au NP, showing Au $4f_{7/2}$ and Au $4f_{5/2}$ peaks at ~83.9 eV and ~87.6 eV, respectively. Bottom: Au 4f spectrum for $Bi_2S_3$@Au, with both peaks found at slightly higher binding energies.

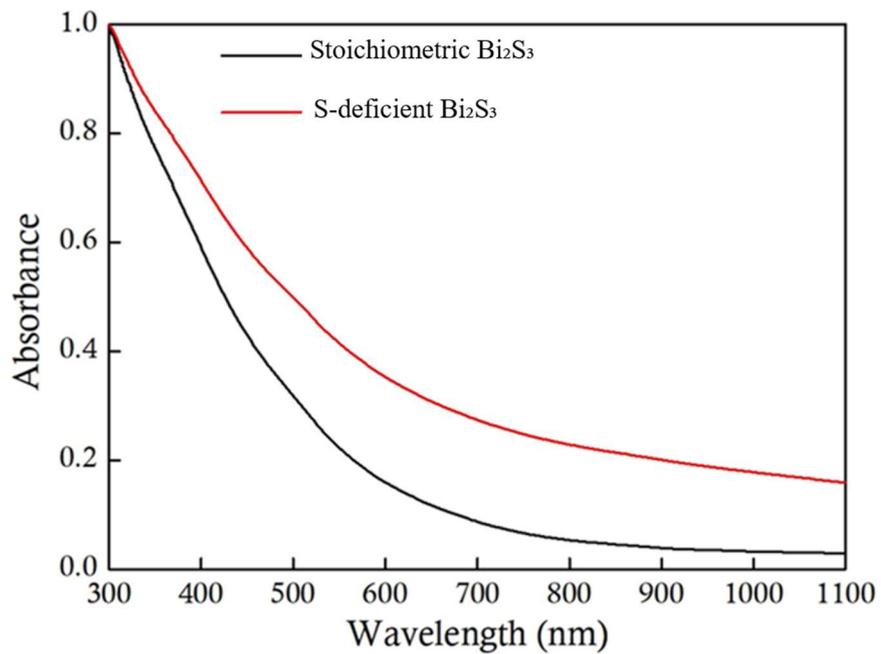

**Figure BS4.** Comparison of UV–vis spectra for $Bi_2S_3$ samples with similar morphology and identical Bi concentration: a stoichiometric $Bi_2S_3$ sample (black) and a S-deficient $Bi_2S_3$ sample (red).

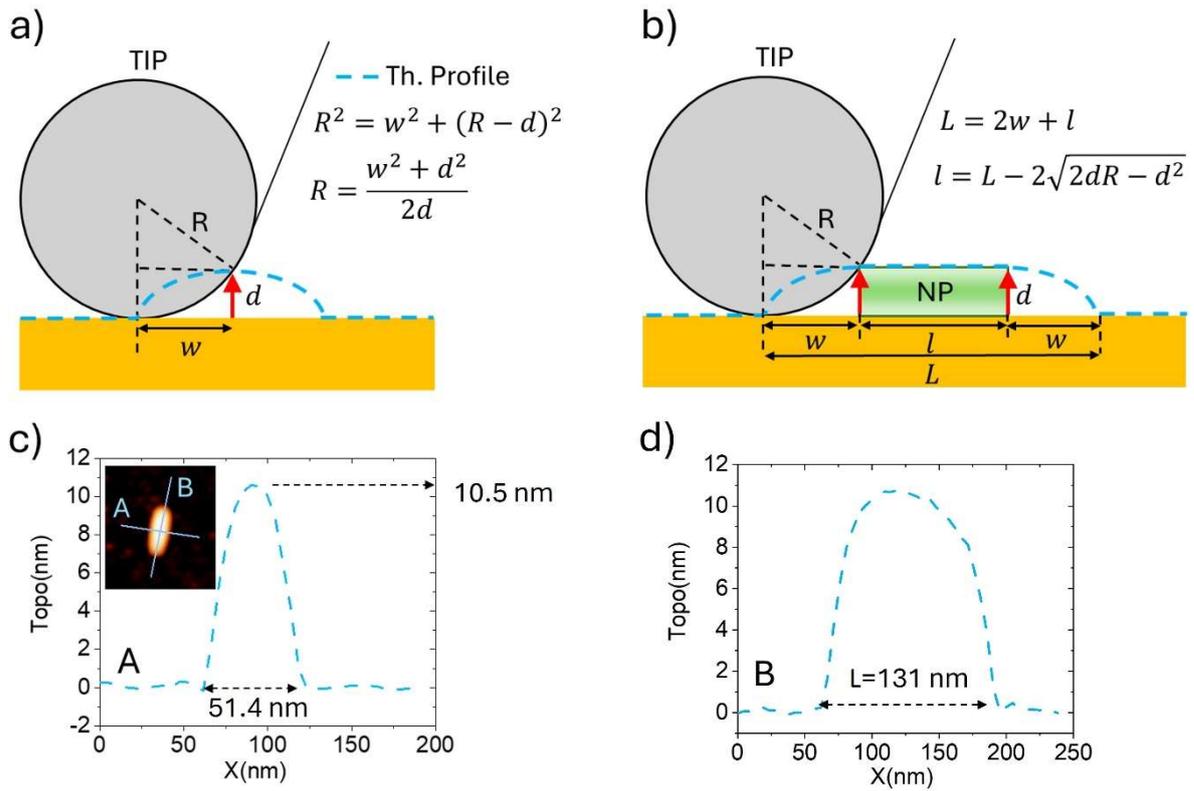

**Figure BS5.** (a), (b) Schematic representation of the tip convolution model used to estimate the real dimensions of NRs from AFM profiles. The diagrams show the geometric relationship between tip radius (R), apparent width (w), and measured height (d), along with the corresponding equations for radius and length correction. (c), (d) AFM topographic profiles of a single $Bi_2S_3$@Au NR taken along directions A and B (as shown in the inset to (c)), with corrected width and length values extracted using the deconvolution model.

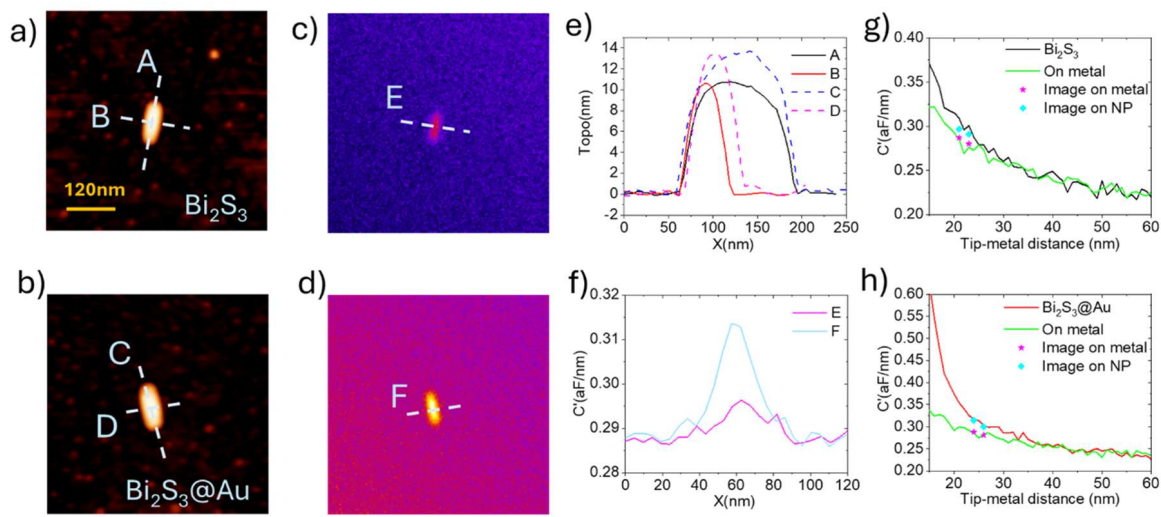

**Figure BS6.** Nanoscale topographic and dielectric characterization of individual $Bi_2S_3$ and $Bi_2S_3$@Au NR. (a), (b) AFM topographic images of $Bi_2S_3$-2 and $Bi_2S_3$@Au-2 NR, respectively. (c), (d) EFM images acquired in constant height mode at tip-sample distances of 23 ± 2 nm for $Bi_2S_3$ and 24 ± 2 nm for $Bi_2S_3$@Au. (e) Corresponding topographic profiles along the dashed lines in (a) and (b): A and B (solid lines) for a $Bi_2S_3$ NR, and C and D (dashed lines) for a $Bi_2S_3$@Au NR. (f) Capacitance gradient profiles (E and F) obtained from (c) and (d), respectively. (g), (h) Electrical approach curves measured over the clean metallic substrate for $Bi_2S_3$ and $Bi_2S_3$@Au samples, respectively.

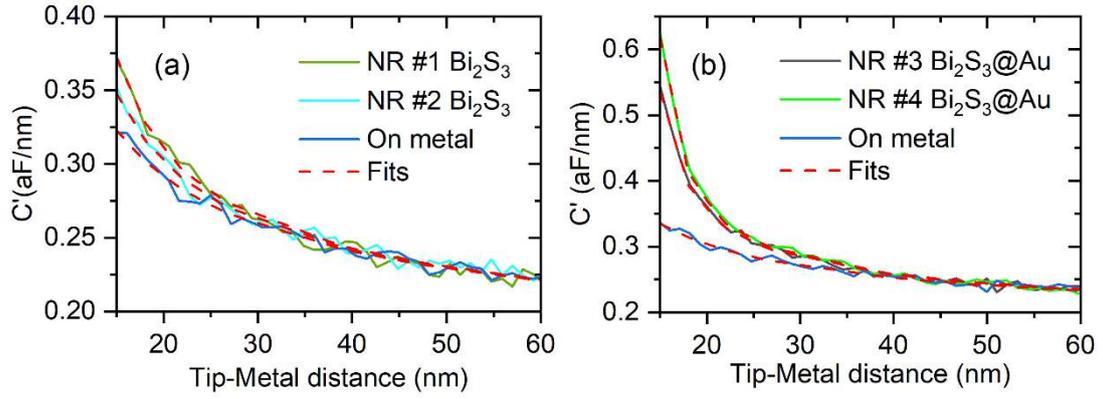

### Table

|  | NR #1 | NR #2 | NR #3 | NR #4 |
|---|---|---|---|---|
| Diameter (nm) | 10.5 | 8.5 | 12.5 | 13.5 |
| Length (nm) | 80 | 75 | 78 | 82 |
| Effective Permittivity | 8±1 | 7±1 | 100±10 | 100±10 |

**Figure BS7.** (a), (b) Approximation curves obtained approximately at the center of the NR and on the metallic substrate. Theoretical fits according to the models described in the main text are shown as red dashed lines. The table summarizes the NR dimensions used in the model, which were obtained from the deconvolution of the topographic images. From the curves acquired on the metallic substrate, we determined the probe parameters and the capacitance offset, C'offset. For the $Bi_2S_3$ sample, we obtained C'offset=89.3±2 zF/nm, while for the $Bi_2S_3$@Au sample the value was 101.4±2 zF/nm. In both cases, the estimation for the tip radius was 35±2 nm and for the cone semi-angle was 22°.

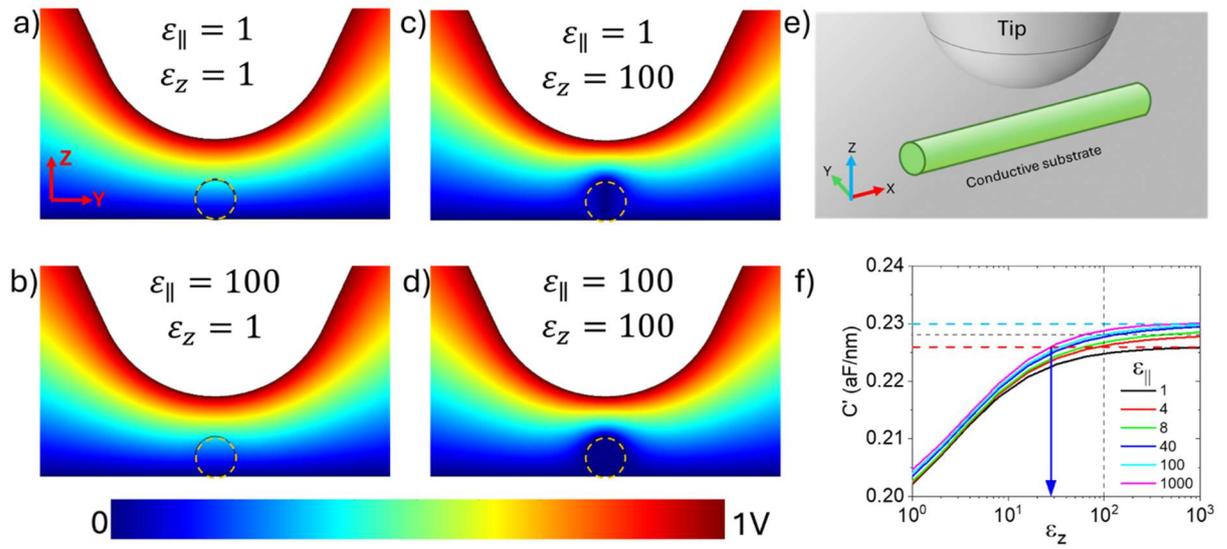

**Figure BS8.** (a)–(d) Simulated electrostatic potential maps in the *y-z* plane across the NR for different anisotropic dielectric constants ($\varepsilon_\parallel$ and $\varepsilon_z$), showing the effect of each component on the potential distribution. (e) Schematic of the 3D model setup with the tip above an anisotropic NR on a conductive substrate. (f) Simulated $C'$ as a function of $\varepsilon_z$ for various fixed values of $\varepsilon_\parallel$, highlighting the system's sensitivity to the perpendicular (*z*-direction) dielectric component. The blue arrow marks the experimental lower bound for $\varepsilon_z$ under the high $\varepsilon_\parallel$ assumption.